\documentclass[twocolumn]{aastex7}
\usepackage{CJK}
\usepackage{color}
\usepackage{amsmath}
\usepackage{ulem}

\newcommand{\D}{{\rm d}}

\begin{document}
\begin{CJK*}{UTF8}{gbsn}
\title{Cosmogenic Neutrino Point Source and KM3-230213A}

\author[0000-0003-1469-208X]{Qinyuan Zhang (张秦源)}
\affiliation{Department of Astronomy, School of Physics, Peking University, Beijing 100871, China}
\email{zhangqy@stu.pku.edu.cn}  


\author{Tian-Qi Huang (黄天奇)}
\affiliation{Key Laboratory of Particle Astrophysics and Experimental Physics Division and Computing Center, Institute of High Energy Physics, Chinese Academy of Sciences, Beijing 100049, China}
\affiliation{TIANFU Cosmic Ray Research Center, Chengdu, Sichuan, China}
\email{huangtq@ihep.ac.cn}

\author{Zhuo Li (黎卓)}
\affiliation{Department of Astronomy, School of Physics, Peking University, Beijing 100871, China}
\affiliation{Kavli Institute for Astronomy and Astrophysics, Peking University, Beijing 100871, China}
\affiliation{TIANFU Cosmic Ray Research Center, Chengdu, Sichuan, China}
\email[show]{zhuo.li@pku.edu.cn}


\begin{abstract}

Cosmogenic neutrinos (CNs) are produced by ultra-high energy cosmic rays (UHECRs) interacting with cosmic background radiation. We investigated the properties of CN point/extended sources, i.e, the neutrino spectrum, and angular profile as functions of time, by assuming that UHECR sources are transient events, such as gamma-ray bursts. The properties depend much on the intergalactic magnetic field (IGMF), but the angular extent is in general sub-degree, within which the CN flux can overshoot the diffuse CN flux in early time. The nearby CN point sources could be detected for the low IGMF case by future neutrino telescopes. The recent KM3-230213A event is possible to account for by a nearby transient CN source, rather than diffuse CN emission. Observations of CN point sources will provide a chance to search for UHECR sources.

\end{abstract}



\section{Introduction} \label{sec:introd}

The origin of ultra-high energy (UHE) cosmic rays (UHECRs; $>10^{19}$eV), is still unknown, but strongly argued to be extragalactic. The propagation of UHECRs in intergalactic medium (IGM) suffers from energy loss due to interaction with cosmic microwave background (CMB) radiation and extragalactic background light (EBL), leading to the Greisen-Zatsepin-Kuzmin (GZK) cutoff \citep{1966PhRvL..16..748G,GZK2}. The suppression of UHECR flux at $\gtrsim5\times10^{19}$eV has been detected by the High Resolution Fly's Eye (HiRes) \citep{HiRes:2007lra}, the Pierre Auger Observatory (PAO) \citep{PierreAuger:2008rol}, and the Telescope Array  \citep{TelescopeArray:2012qqu}, consistent with the GZK cutoff of UHECR protons. Moreover, the small Milky Way magnetic field ($\sim\mu$G) can not confine UHECRs, and the observed anisotropy of their arrival direction does not support Milky Way but extragalactic origin \citep{PierreAuger:2017pzq}.  

It was predicted that a guaranteed diffuse flux of UHE neutrinos, along with the GZK cutoff, is produced when UHECRs propagate in IGM \citep{1969PhLB...28..423B,1979ApJ...228..919S}, i.e., the so-called cosmogenic neutrinos (CNs). UHE neutrino detection will be the goal of many future neutrino telescopes, such as IceCube-gen2 \citep{IceCube-Gen2:2020qha} and GRAND \citep{GRAND:2018iaj}. Theoretical works suggest that the observation of CNs will provide information on the spectrum and production rate of UHECRs, as well as the cosmological evolution of UHECR sources \citep{2001PhRvD..64i3010E,2006JCAP...09..005A, Kotera:2010yn,PhysRevD.95.063010}. The latest observation of diffuse CN emission by IceCube \citep{IceCube:2018fhm} and PAO \citep{PierreAuger:2023pjg} put upper limits on the flux close to the predicted upper bound of UHE neutrino emission, i.e. the Waxman-Bahcall bound \citep{wbb}, and already put meaningful constraints on some models of UHECR sources.

In fact, the all-sky diffuse CN flux should be contributed by individual sources. The propagation length of UHECRs is limited by the energy loss, say a few hundreds Mpc for protons of $10^{20}$eV. Thus, for a distant observer the CN emission appears as a single point source/extended source. Observation of CN point sources (CNPSs) may provide a great opportunity to solve the UHECR origin problem by searching for the CNPS-associated astrophysical counterparts. 
For the case of steady and spherically symmetric nearby UHECR sources, see \cite{PhysRevD.82.043002} which studies the spectrum and angular distribution of UHE protons and secondary particles of individual sources, taking into account the scattering of UHECRs by intergalactic magnetic field (IGMF) in propagation (see also studies on secondaries from distant steady blazars in the case of very low IGMF, e.g., \cite{2010PhRvL.104n1102E,2012ApJ...749...63M,2022A&A...658L...6D}).

However, UHECR sources are likely powerful astrophysical transients, usually with relativistic jets, such as bright flares from blazars or gamma-ray bursts (GRBs; see, e.g., \cite{2011ARA&A..49..119K}). The CN sources induced by transient and jetted astrophysical events should be considered. 
Here we take GRBs as an example to investigate the spectrum and size of a CNPS, as well as its temporal evolution, resulting from UHECRs produced by a transient. We also discuss the possibility that the recent reported UHE neutrino event KM3-230213A (KM3 event, hereafter) 
\citep{KM3NeT:2025npi} originates from a nearby transient CNPS. 

\section{Model} \label{sec:model}
As shown in \autoref{fig:EATS}, we establish a spherical coordinate system $(r, \theta, \varphi)$ centered at the origin ($r=0$), with the UHECR source positioned at this central point.
Consider that a GRB releases a beam of UHECRs into the IGM toward the Earth at time $t=0$. 
Denote that the GRB distance is $d$, and the opening angle of the UHECR beam is equal to that of the GRB jet, $\theta_{\rm j}$\footnote{Because of the ultrarelativistic motion of the GRB jet, typically with a Lorentz factor $\Gamma>100$, UHECRs escape with a negligible angle $<1/\Gamma$ with respect to the radial direction, even if isotropically disributed in the rest frame of the jet.}. If the transient duration is much shorter than all timescales relevant to CN production, we can effectively consider that all the UHECRs are released instantaneously. 
We will assume that UHECRs are protons. The initial distribution of CR protons is a flat power law\footnote{A flat initial spectrum is required to accommodate the observed UHECR spectrum to the GZK effect \citep[e.g.,][]{2009JCAP...03..020K}.} with exponential cutoff, $N_p(E_p)=dN_p/dE_p=KE_p^{-2}\exp(-E_p/E_{\rm cut})$, where $E_{\rm cut}=3\times10^{20}$eV is set to be the highest CR energy ever detected \citep{1995ApJ...441..144B}. 
The coefficient $K$ is determined by assuming the total isotropic-equivalent CR energy $\int_{E_{\min}} E_p N_p \D E_p\simeq20KE_p^2 =f_pE_{\rm iso}$, with $f_p$ the CR loading factor and $E_{\rm iso}$ the isotropic-equivalent GRB energy in gamma rays.

\subsection{Propagation}\label{sec:model_IGMF}
The propagation of CRs suffers from IGMF scatterings. 
Due to the high proton energy and small magnetic strength, we can adopt the small-angle multiple scattering approximation for the propagation of protons in the IGMF, which is simply assumed to be isotropic and homogeneous \citep[e.g.,][and references therein]{PhysRevD.82.043002}.
The mean square scattering angle of a proton per unit propagation length is given by 
\begin{equation}
    a_{\rm s}(E_p)\equiv\langle \theta_{\rm s}^{2} \rangle
    =(\lambda/5) (e/E_{p})^{2} \langle \mathbf{B}^{2}\rangle,
\end{equation}
where $\lambda$ and $\mathbf{B}$ are the IGMF correlation length and strength, respectively, and $E_p$ is the proton energy \citep{PhysRevD.82.043002}. After propagating a distance $r$, the characteristic deflection angle of the proton with energy $E_p$ is $\theta_{c} = \sqrt{a_{\rm s} r}$, i.e.,
\begin{equation}\label{eq:theta_c}
    \theta_{c} = 0.13 \left(\frac{\lambda}{1{\rm Mpc}}\right)^{\frac12} \frac{B}{1{\rm nG}}\left(\frac{ E_{p}}{10^{2} {\rm EeV}}\right)^{-1} \left(\frac{r}{1{\rm Gpc}}\right)^{\frac12}
\end{equation}
with $B=\langle \mathbf{B}^{2}\rangle^{1/2}$ the root mean square of IGMF.
For a wide range of parameter values, $\theta_{\rm c}\ll1$, thus we adopt the small-angle approximation for proton propagation in the following.

The characteristic radial propagation velocity of protons can be given by 
\begin{equation}
    v_{\rm c}=c\cos\theta_{\rm c}\simeq c (1 - \theta_{\rm c}^2/2),
\end{equation} 
where the last equation is for the small-angle approximation.  The typical time it takes for a proton of energy $E_p$ to propagate to radius $r$, $t=\int_0^r\D r'/v_{\rm c}(r')$, is
\begin{equation}\label{eq:r_t}
    t \simeq \frac{2}{ca_{\rm s}} \ln \left[\left(1-\frac{a_{\rm s} r}{2} \right)^{-1}\right] \simeq \frac{r}{c}+\frac{a_{\rm s} r^2}{4 c},
\end{equation}
where the second term results from expansion to the second order of $a_{\rm s} r$, and is the additional propagation time due to IGMF scattering. 

\begin{figure}[t] 
    \centering
    \includegraphics[width=\linewidth]{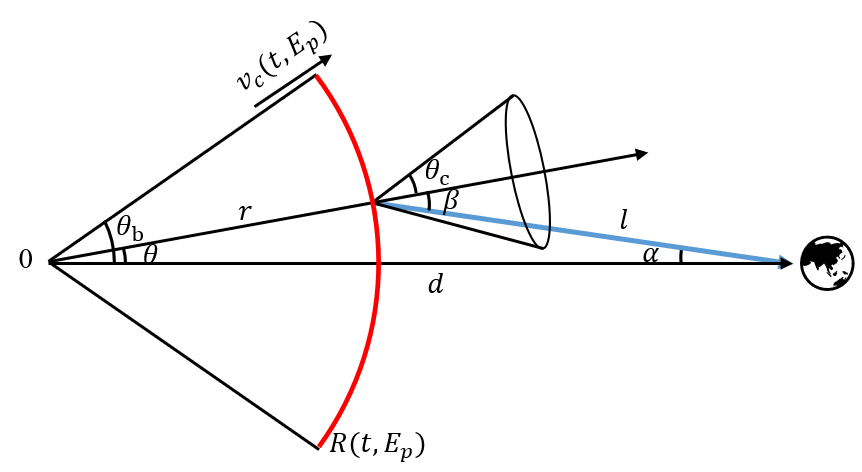}
    \caption{Schematic plot of propagating protons of certain energy, forming a part of an expanding spherical surface (red arc), and emitting neutrinos (blue line).}
    \label{fig:EATS}
\end{figure}

Notice that $\theta_{\rm c}$ and hence $v_{\rm c}$ depend on $E_p$. We take a simple picture for proton propagation, as shown in Figure \ref{fig:EATS}: 
\begin{itemize}
    \item At any time $t$, protons of the same energy $E_p$ are assumed to be located on a spherical surface of radius $R=r(t,E_p)$ (derived from \autoref{eq:r_t}); 
    \item The proton number density on the sphere is uniform; 
    \item The opening angle of the proton beam increases with $R$, due to IGMF scattering,  $\theta_{\rm b}\simeq\max(\theta_{\rm j},\,\theta_{\rm c})$; 
    \item In a surface element of the sphere, the angular distribution of the velocity vector of protons is approximated as a uniform distribution in the cone of opening angle $\theta_{\rm c}$, with the central axis of the cone along the normal of the surface element. 
\end{itemize}
We adopt Minkowski spacetime, as we are concerned about sources in the local universe $d\lesssim100$ Mpc.

\subsection{Energy loss and neutrino production}
Protons lose energy by $p\gamma$ interactions during propagation. The proton number at $R$ for proton energy of $E_p$ to $E_p+\D E_p$ is approximated as 
\begin{equation}
    N_p(R)\D E_p=N_p(0)e^{-R/d_{\rm eff}}\D E_p,
\end{equation} 
where $d_{\rm eff}(E_p)=ct_{\rm eff}(E_p)$ is the effective energy loss length. We estimate the effective energy loss time of UHECR protons, $t_{\rm eff}$,  using the parameterized description from \cite{2009JCAP...03..020K}, which takes into account both pair production and photopion production in $p\gamma$ interactions, 
\begin{equation}
    t_{\rm eff}^{-1}(E_p) = t_{0,ep}^{-1} e^{-E_{c,ep}/E_p} + t_{0,\pi}^{-1} e^{-E_{c,\pi}/E_p},
\end{equation} 
with $E_{c,ep} = 2.7 \times 10^{18} \rm \, eV$, $t_{0,ep} = 3.4 \times 10^{9} \rm \, yr$, $E_{c,\pi} = 3.2 \times 10^{20} \rm \, eV$, and $t_{0,\pi} = 2.2 \times 10^{7} \rm \, yr$. 
Because of the increase of the opening angle of the CR beam relative to the initial GRB jet, the proton number per unit proton energy per unit solid angle of the sphere at $r=R$ should be
\begin{equation}
    \frac{\D N_p(R)}{\D \Omega_{\rm e}} =\frac1{4\pi}N_p(0)e^{-R/d_{\rm eff}(E_p)}\frac{\theta_{\rm j}^2}{\theta_{\rm b}^2(R,E_p)},
\end{equation}  
where $\D\Omega_{\rm e}=2\pi\sin\theta\D\theta$.


For $p\gamma$ interactions and neutrino production, we use the parameterization method of \cite{2008PhRvD..78c4013K}. If a proton of energy $E_p$ interacts with isotropic background photons, where the photon number density per unit photon energy $\epsilon$ is $f_{\rm ph}(\epsilon)$, the all-flavor neutrino number produced per unit time $t$ per unit neutrino energy $E_\nu$ can be given by 
\begin{equation}\label{eq:Kelner2008}
    Q_\nu(E_{\nu},E_p) = \frac{1}{E_p} \int f_{\rm ph}(\epsilon) \Phi_\nu(\eta,x) \D \epsilon,
\end{equation}
where $\eta \equiv 4\epsilon E_{p}/(m_p c^2)^2$, $x \equiv E_{\nu}/E_p$, and $\Phi_\nu(\eta,x)$ presents the neutrino spectrum produced by interactions with fixed $E_p$ and $\epsilon$. For each neutrino type $f$, the neutrino spectrum $\Phi_f$ is different, and the parametrized forms of function $\Phi_f$ have been given in section II.B of \cite{2008PhRvD..78c4013K}. Adopting the forms for $f=\nu_\mu,\,\bar{\nu}_\mu,\,\nu_e$,  we obtain the total all-flavor neutrino spectrum from pion decay
$    \Phi_{\nu}(\eta, x) =\sum_f\Phi_f(\eta, x)$.
In addition to CMB, we also consider EBL adopted from \cite{2017A&A...603A..34F} for $f_{\rm ph}(\epsilon)$.

Denote $\tau$ the observer time; and $\tau=0$ corresponds to the point at which a photon emitted at $r=0$ at $t=0$ arrives at Earth. The observed neutrino number intensity per unity proton energy at viewing angle $\alpha$ (Figure \ref{fig:EATS}) is given by
\begin{equation}
    \frac{\D I_{\nu}}{\D E_p}
    = Q_{\nu} \frac{\D N_p(R)}{\D \Omega_{\rm e}}\frac{\Theta ( \theta_{c} - \beta )}{\Delta\Omega_{\rm c}l^2}\frac{\partial t}{\partial\tau}.
\end{equation}    
Here $\beta=\theta+\alpha$ is the angle of the light of sight with respect to the normal of the surface element, $\Theta$ is the Heaviside function accounting for that beyond $\theta_{\rm c}$ neutrino emission is negligible, $\Delta\Omega_{\rm c}(R)\approx\pi\theta_{\rm c}^2$ is the solid angle of the proton beam and hence the cone of emitted neutrinos, as neutrinos are in fact emitted along the direction of the primary protons \footnote{The deflection of secondary muons and pions before decay is negligible \citep{LZ07}.}, $l$ is the distance between the emitting element and the Earth, and $\partial t/\partial \tau$ is the relation between the coordinate time and the observer time for fixed $\theta$ and $E_p$ (considered below).

\subsection{Neutrino flux and angular profile}
The observer time $\tau$ is the time difference between the observed arrival of neutrinos emitted at location $(r,\theta)$ and that of photons emitted at $r=0$ when $t=0$, thus, $\tau = t + (l/c) - (d/c)$). Plugging $t(r,E_p)$ (\autoref{eq:r_t}) and $l(r,\theta) = \sqrt{d^2 + r^2 - 2rd\cos\theta}$ one obtains
\begin{equation}\label{eq:tau}
    c \tau=\frac{a_{\rm s}(E_p) r^{2}}{4}  +\sqrt{d^2 + r^2 -2rd\cos\theta} -(d-r). 
\end{equation}
Once $\tau$ and $E_p$ are fixed, one obtains a $r$-$\theta$ relation, which defines an equal arrival time surface (EATS) for protons of energy $E_p$, $R_\tau=r(\theta;\tau, E_p)$. The neutrinos are emitted from the EATS at different $t$ but arrive at Earth at the same $\tau$. 

At $\theta=0$ one has $R_{\tau,0}=2\sqrt{c\tau/a_{\rm s}}$ (for $R_{\tau,0}<d$). The EATS shape is much elongated. By small-angle approximation, one can derive that the EATS radius as function of $\theta$ is
\begin{equation}\label{eq:Rtau}
    R_\tau\simeq\left\{\begin{array}{ll}
        R_{\tau,0} & \theta\lesssim\theta_{\rm w},\\
        2c\tau/\theta^2 & \theta\gtrsim\theta_{\rm w},
    \end{array} 
    \right.
\end{equation}
where $\theta_{\rm w}\equiv(c\tau a_{\rm s})^{1/4}\sim\theta_{\rm c}(\theta=0)$, beyond which the EATS radius decreases rapidly with $\theta$.
With the EATS shape and evolution solved above, one can derive, for a certain $\theta$, $\partial t / \partial\tau = (\partial R_\tau/v_{\rm c} \partial\tau)_\theta$, required for the calculation of intensity.

The EATS depends on $E_p$ and $\tau$. The total observed neutrino flux at time $\tau$ should be integration over EATS and over proton energy distribution, which is
\begin{equation}
    \phi_{\nu}(E_{\nu},\tau) =
    \int \D E_p \int_{\{R_\tau\}} \D \Omega \frac{\D I_{\nu}}{\D E_p }\cos\alpha,
\end{equation}     
where $\D\Omega=2\pi\sin\alpha\D\alpha$ is the observing solid angle, and the  inner integration should be done over the EATS, denoted as $\{R_\tau\}$, for certain $E_p$ and $\tau$.

Since the CN source may be extended in observations, it is useful to consider the size and angular distribution of the source. Given the above result, we can calculate the neutrino intensity as a function of $\alpha$ for the CN source at time $\tau$, $I_{\nu}(E_{\nu},\tau,\alpha) = \D \phi_{\nu}(E_{\nu},\tau,\alpha) / \cos\alpha \D\Omega$, and obtain the angular size.

\section{Result} \label{sec:results}
The typical GRB emission energy and jet opening angle are derived to be $E_{\rm iso}\sim10^{53}$erg, and $\theta_j\sim0.1$ \citep[See e.g.,][for reviews]{2002ARA&A..40..137M,2015PhR...561....1K}. For GRBs to be the UHECR sources, the  typical loading factor should be $f_p\sim10$ \citep[e.g.,][]{1995PhRvL..75..386W,2004ApJ...606..988W}, which is also consistent with the constraint by IceCube \citep{2017ApJ...843..112A}.
We adopt the following fiducial values, 
unless otherwise stated: $d = 100$ Mpc, $E_{\rm iso} = 10^{53}$erg, $f_{p} = 10$, $\theta_{\rm j}=5^{\circ}$, and $\lambda= 1$ Mpc. The IGMF is largely unknown, with a wide range allowed by current observational constraints; it can be up to $B\lesssim10^{-9}$G \citep{Planck:2015zrl,2016PhRvL.116s1302P} or down to $B\gtrsim10^{-16}$G \citep{2010Sci...328...73N}. We consider two cases: $B = 10^{-12}$G (denoted as M1 case) and $10^{-9}$G (M2). 

For both M1 and M2, we have $\theta_{\rm c}\lesssim\theta_{\rm j}$ (\autoref{eq:theta_c}), and then the proton beam opens an angle of $\theta_{\rm b}\simeq\theta_{\rm j}$. Since only neutrino emission from $\theta<\theta_{\rm c}$ can be observed, the neutrino flux is not limited by the jetted geometry in this case, thus the CN source will appear the same as an emitting sphere.

It is useful to calculate two characteristic observer times. The first is the observer time that protons with $E_p$ propagate a distance $d$ (forgetting energy loss).
Taking $\theta=0$ and $r=d$ in \autoref{eq:tau}, the observer time is estimated as $\tau_{\rm d}  \simeq d^2 a_{\rm s}/ 4c$, i.e.
\begin{equation}\label{eq:tau_d}
    \tau_{\rm d}  \simeq 0.13 \left(\frac{E_p}{10^{20} {\rm eV}}\right)^{-2} \left(\frac{B}{10^{-12}{\rm G}}\right)^2 \left(\frac{d}{10^2{\rm Mpc}}\right)^2 {\rm yr}.
\end{equation}
The other is the observer time that protons propagate a distance $d_{\rm eff}$ ($d>d_{\rm eff})$, $\tau_{\rm eff}  \simeq d_{\rm eff}^2 a_{\rm s}/ 4c$, which can be estimated by replacing $d$ with $d_{\rm eff}(E_p)$ in \autoref{eq:tau_d}.

M1 and M2 are two typical cases. In M1, the observed propagation time $\tau_{\rm d}$ and energy loss time $\tau_{\rm eff}$ are small. The energy loss length of protons with $E_p=100$ EeV is $d_{\rm eff}\sim100$ Mpc. Thus protons with $E_p\lesssim100$ EeV arrive and pass through Earth in a short observer time. However, for M2, the observed propagation time to $d$ is large, $\tau_{\rm d}\gtrsim10^5$yr for $E_p\lesssim100$ EeV.

\begin{figure}[t]
    \centering
    \includegraphics[width=\linewidth]{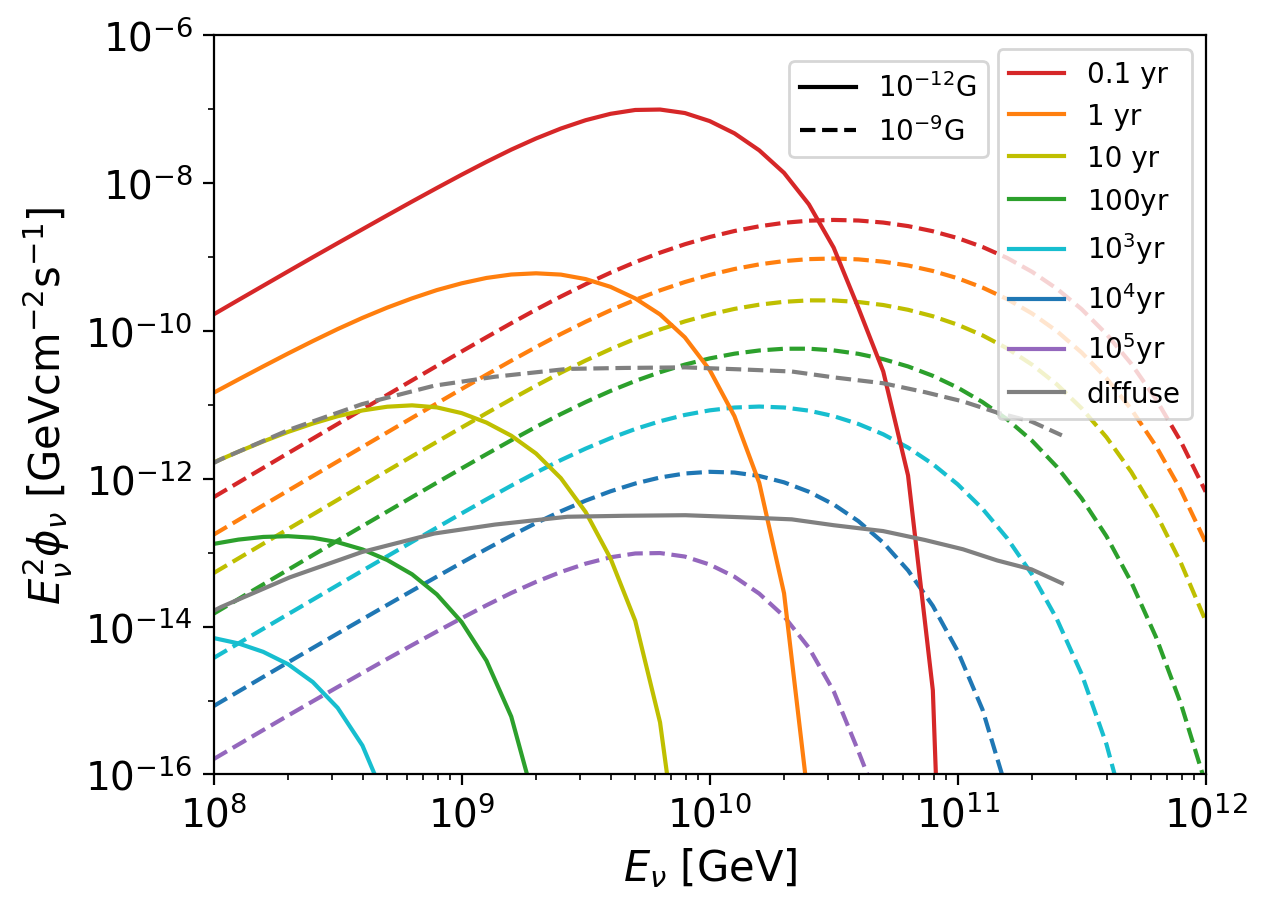}
    \caption{Temporal evolution of all-flavor neutrino spectrum for $d=100$ Mpc. The colors represent different observer times $\tau$, as marked. The solid and dashed lines correspond to M1 and M2, respectively.  The gray solid and dashed line present the diffuse CN flux \citep[from][]{2001PhRvD..64i3010E} from a region within $0.1^{\circ}$ and $1^\circ$, respectively.}
    \label{fig:temp_spectrum}
\end{figure}

\subsection{Neutrino spectrum}
The numerical results of neutrino spectra at different observer time are presented in \autoref{fig:temp_spectrum} for both M1 and M2 cases. The behaviors are very different in between. 
In M2, at time $\tau<\tau_{\rm eff}(E_{\rm cut})\sim400$ yr ($d_{\rm eff}(E_{\rm cut})\sim20$ Mpc), the spectral profile does not change significantly, but the flux decreases slowly over time. This can be understood as follows. First of all, all protons ($E_p<E_{\rm cut}$) do not suffer significant energy loss (since the energy loss length is larger for lower-energy protons), so the proton spectrum, and hence the neutrino spectrum, remains unchanged. The observable solid angle of the part of the sphere increases as $\propto\theta_{\rm c}^2$, but the neutrino intensity decreases as $\propto\theta_{\rm c}^{-2}$, so the neutrino flux scales only as $\phi_\nu\propto\partial t/ \partial\tau \simeq \partial R_\tau/c\partial\tau\simeq (c\tau a_{\rm s})^{-1/2}$, taking $R_\tau\simeq R_{\tau,0}$ in the last equation. Thus, the flux only decreases as $\phi_{\nu} \propto B^{-1} \tau^{-1/2}$ for $\tau<\tau_{\rm eff}(E_{\rm cut})$.  
At $\tau>\tau_{\rm eff}(E_{\rm cut})$, lower-energy protons start to lose energy significantly, with the proton energy determined by $\tau=\tau_{\rm eff}(E_p)$. This leads to a proton spectral turnover and hence the peak energy of the neutrino spectrum moves to lower energies with time, as seen in \autoref{fig:temp_spectrum}.

In M1 the spectrum evolves faster. Since $\tau_{\rm eff}(E_{\rm cut})\sim5$ hr is small, the peak energy of the neutrino spectrum starts to decrease much earlier than that in M2, as seen in \autoref{fig:temp_spectrum}. It also can be seen that the spectrum moves quickly to low energy end. As $\tau_{\rm d}$ is small, CRs can reach Earth in a short observer time. Since $\tau_{\rm d}\propto E_p^{-2}$, lower energy protons arrive later. After arrival and pass through, the neutrino flux at the related energy, $E_\nu\sim0.01E_p$, will drop abruptly, since the emitted neutrinos are beamed along the radial direction. Consequently the neutrino spectral cutoff evolves quickly toward low energies.

In \autoref{fig:temp_spectrum}, it can be found that the spectrum of M1 ($B=10^{-12}$G) at $\tau=10^{-1}$yr shows a spectral shape similar to that of M2 ($B=10^{-9}$G) at $\tau=10^{5}$yr, but the flux is larger by a factor of $\sim10^6$, comparable to the difference between two times $\tau$'s, that is, the ``fluence" $\phi_\nu\tau$ is roughly constant between them. It can be understood as follows. For cases with the same $\tau/B^2$, \autoref{eq:r_t} implies that (for $\theta=0$) source-frame time $t$ is similar, thus the protons propagate to a similar radius and lose similar energy, resulting in similar proton energy distribution and turnover energy and hence neutrino spectral profile. Since protons lose and emit similar amount of energy within time $t$, $\phi_\nu\tau\sim$ constant.

\begin{figure}[t]
    \centering
    \includegraphics[width=\linewidth]{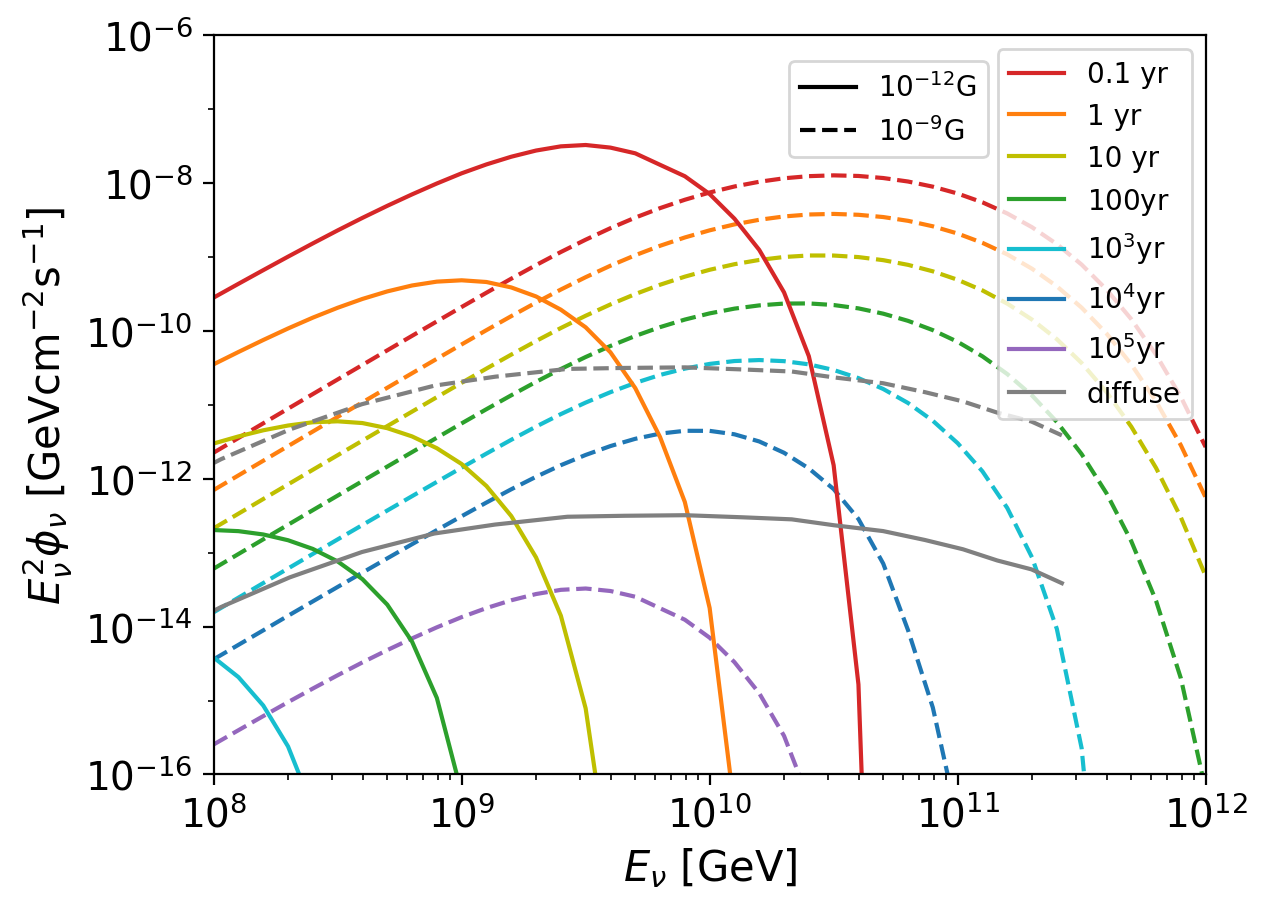}
    \caption{Same as \autoref{fig:temp_spectrum}, but for $d=50$ Mpc.}
    \label{fig:temp_spectrum_50Mpc}
\end{figure}

For comparison we also show in \autoref{fig:temp_spectrum_50Mpc} the case of $d=50$ Mpc. Compared with the $d=100$ Mpc case, the neutrino flux increases across the entire observation time at energies below the spectral break, and the break energy is lower, especially for M1 case. This is caused by proton propagation effect. The propagation time depends on distance and energy as $\tau_{\rm d}\propto d^{2} E_{p}^{-2}$, so that for given time $\tau_{\rm d}$ the break energy is proportional to $d$.

\begin{figure}[t]
    \centering
    \includegraphics[width=\linewidth]{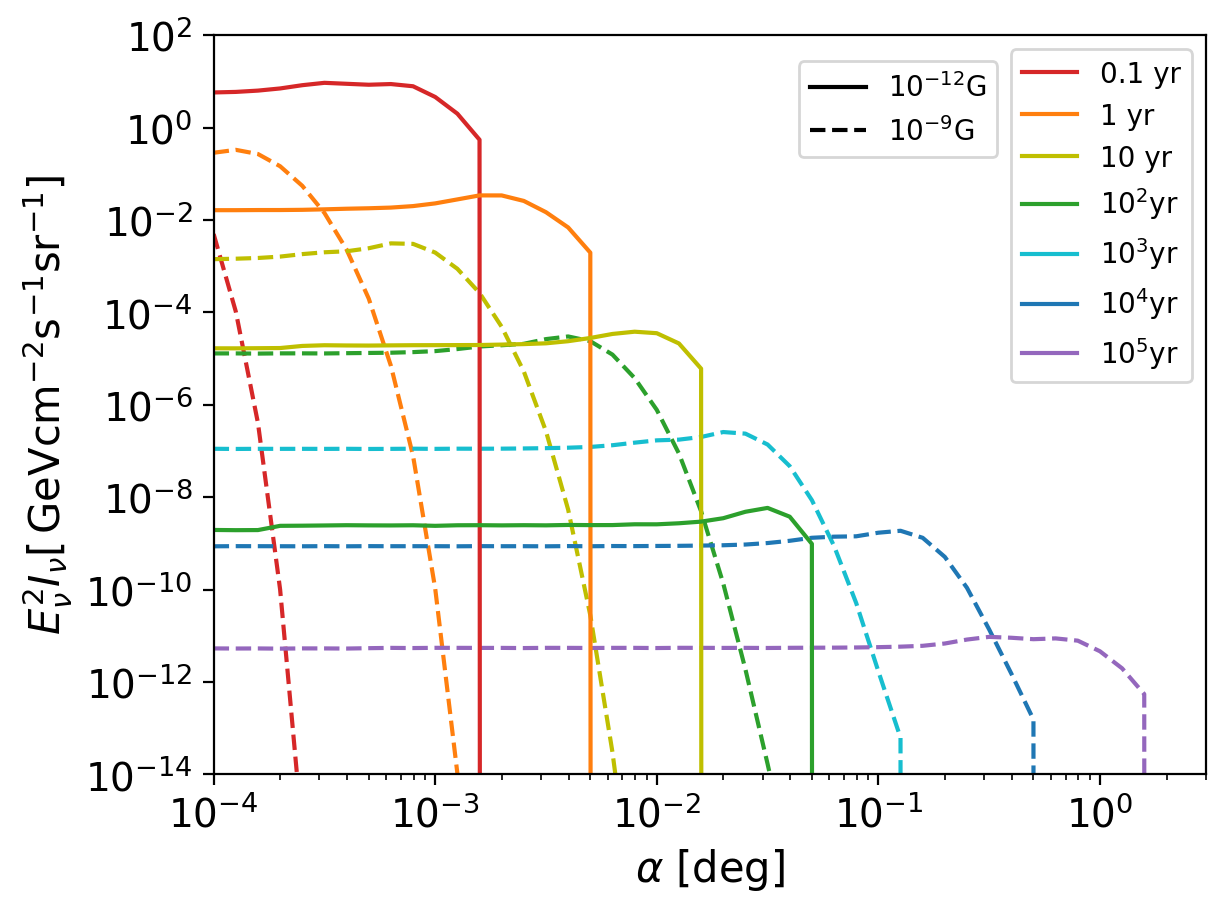}
    \caption{Neutrino intensity at $E_\nu=1$ EeV as function of viewing angle $\alpha$ for $d=100$ Mpc. The solid and dashed lines correspond to M1 and M2, respectively. The colors represent different observer times $\tau$. }
    \label{fig:angular_profile}
\end{figure}

\subsection{Neutrino angular profile}
As for the angular profile, \autoref{fig:angular_profile} shows the calculated observed neutrino intensity at $E_{\nu} = 1$\,EeV as a function of observing angle. In general, the neutrino intensity is distributed uniformly with angles, but a slight limb brightening is seen, which is caused by the EATS effect for relativistically expanding but emission-decaying spheres. 
That is, in the EATS $R_\tau$ decreases with increasing $\theta$, as a result, protons at small $\theta$ suffer more significant energy loss, leading to a dim center with a brighter limb.

From \autoref{fig:angular_profile}, the angular sizes and profiles are also very different between M1 and M2: M1 has larger size; M2 shows the size increases faster with $\tau$; The angular extent can be larger than $1^\circ$ at late times in M2, but never reach $0.1^\circ$ in M1. We discuss the cause here. Neutrino emission at 1 EeV is dominated by protons of $\gtrsim10^2$EeV, if the suppression of protons by energy loss is not important yet. In M2, if $\tau<\tau_{\rm eff}(10^2\rm EeV)$, the observed neutrino flux at $E_\nu\sim1$ EeV remains unaffected. Because of small angular spreading of protons, neutrino emission is mainly contributed by the part of the EATS with $\theta \lesssim \theta_{\rm c}(10^2\rm EeV)$. At such small angles $\theta$, the EATS radius is $R_\tau(\theta)\simeq R_{\tau,0}$ (\autoref{eq:Rtau}). The angular size  can be estimated by the spreading of protons with $E_p\sim10^2$EeV,
$\alpha_{\max} \simeq \theta_{\rm c} R_{\tau,0}/ d = 2 (c\tau)^{3/4}a_{\rm s}^{-1/4}d^{-1}$, i.e., $$\simeq 2\times10^{-4} (\tau/1 \,{\rm yr})^{3/4} (E_p/10^2{\rm EeV})^{1/2}(B/1\,\rm nG)^{-1/2}\rm degree $$
for $\tau \lesssim 10^3$yr in M2.

\autoref{fig:angular_profile} shows sharp edges of angular profiles for M1 (and for M2 with large $\tau$, i.e., $\tau\sim10^5$yr, as well). 
The abrupt suppression at the edge is caused by the fact that the highest energy CRs reaching the Earth do not contribute to the observed neutrinos any more.
Note that higher energy protons have larger angular extent, $\propto R_{\tau,0}\theta_{\rm c}\propto a_{\rm s}^{-1/4}\propto E_p^{1/2}$. In M2, the observed neutrinos emission at larger $\alpha$ with lower intensity is contributed by protons of $E_p\gtrsim10^2E_\nu$, which suffer stronger suppression due to energy loss. Once these protons reach $R_{\tau,0}(E_{p})\simeq d$, the wing emission of large angles is suppressed and hence the edge becomes sharper.
One can derive, by \autoref{eq:tau_d}, that the energy of the protons that are reaching is $E_{p,\rm d} \simeq 37 (B/10^{-12}{\rm G})(\tau/1 {\rm yr})^{-1/2}  {\rm EeV}$. In the case of $E_p\lesssim10^2$EeV, the angular extent of the edge can be roughly estimated by $\alpha_{\max} \simeq R_{\tau,0}(E_{p,\rm d})\theta_{c}(E_{p,\rm d})/d$, i.e., $$\simeq 4\times10^{-3}(\tau/1 \,{\rm yr})^{1/2} (B/10^{-12} {\rm G})^{-1/4}\rm degree,$$
which is consistent with the time evolution of the sharp edge in M1. Finally, at $\tau = 10^3$ yr, we have $R_{\tau,0}(E_p)\sim 100 \rm Mpc$ for $E_p=1$ EeV, thus, no neutrino flux is observed at $E_\nu>1$ EeV any more as protons of $E_p>1$ EeV pass through the Earth. Then the evolution of the angular size for $E_\nu=1$ EeV stops with small angles of $<0.1^\circ$ in M1 (\autoref{fig:angular_profile}).

\section{KM3-230213A} \label{sec:KM3evt}

The KM3 event \citep{KM3NeT:2025npi} is a UHE muon traversing the KM3NeT-ARCA array. The reconstructed direction is $\rm RA = 94.3^{\circ}$, $\rm dec. = -7.8^\circ$, 
with the $68\%$ uncertainty of $1.5^{\circ}$. The reconstructed muon energy is $120^{+110}_{-60}\,{\rm PeV}$. Here, we consider the possibility that the KM3 event originates from a CNPS, in contrast to diffuse CN emission\footnote{See also the other possibilities, e.g., the decay of superheavy dark matter \citep{2025arXiv250322465C,2025arXiv250307776N,2025arXiv250304464K}.}.

With the effective area of a neutrino telescope, $A_{\rm eff}(E_{\nu})$, the number of neutrino events detected in a period of time $T$ within an energy range of $(E_1,E_2)$ can be calculated as $N_\nu= \int_0^T\int_{E_1}^{E_2} A_{\rm eff}(E_{\nu})\phi_{\nu}(E_{\nu},\tau)\D E_{\nu}\D \tau$. The KM3  event could be induced by isotropic diffuse neutrinos or by a neutrino point source in the arrival direction. The effective areas of telescopes for isotropic neutrinos and neutrino point sources in the KM3 event direction (with declination $\delta_s=-7.8^{\circ}$) are compared in \autoref{fig:Aeff} in Appendix \ref{sec:effective area}. For isotropic sources, we take the full-sky averaged effective areas for IceCube EHE events \citep{Meier:2024flg} and for KM3NeT-ARCA real-time alerts (21-line configuration; hereafter ARCA21)  \citep{KM3NeT:2025npi}. For point sources, the effective area of IceCube is taken from \cite{IceCube:2021xar}, while for ARCA21 we derive it from the upgoing effective area \citep{2024icrc.confE1018M}, scaled by an energy-dependent factor $\xi(E_{\nu})$ (\autoref{eq:xi})
to account for the zenith distribution of events in the point source's direction.

If the KM3 event originates from diffuse CNs, assuming the diffuse CN flux from \cite{2001PhRvD..64i3010E} and taking the full-sky averaged effective areas of ARCA21 and IceCube, the muon neutrino events (assumed to be one third of the all-flavor neutrino number after neutrino mixing in propagation) detected in the energy range of 120 PeV to 3 EeV are expected to be $N_\nu\simeq0.0028$ for $T\simeq1$ yr by ARCA21, while $N_\nu\simeq0.19$ for $T\simeq 10 $ yr by IceCube (The different $T$'s reflect different operation times between two telescopes). The large contrast of a factor of $\sim68$ is in tension with the non-detection of UHE neutrinos by IceCube but one detection by ARCA21. However, if the KM3 event originates from a nearby transient CNPS, in M1 case the expected muon neutrino events detected within $T\simeq1$ yr after a transient event (for example, a GRB) are $N_\nu\simeq0.0012$ by ARCA21, and $N_\nu\simeq0.017$ by IceCube. The contrast in between reduces to a factor of $\sim14$, much smaller than that of diffuse CN origin. So the transient CNPS origin can relieve the tension between ARCA21 and IceCube compared with the diffuse CN emission, and hence a transient CNPS is preferred over diffuse CN emission for the KM3 event\footnote{The tension between ARCA21 and IceCube implies a transient origin rather than steady sources \citep{2025arXiv250212986N}.}.

However, as IceCube is $\sim10$ times more sensitive than ARCA21 in the KM3 event direction (see the contrast between green and blue lines in \autoref{fig:Aeff}), always more events were expected to be detected by IceCube than ARAC21. Thus, the tension indicates that the expected event number should be low for both telescopes, and the detection of the KM3 event should rely on fluctuations in low statistics. But, to avoid too low statistics, one still needs nearby energetic GRBs. For examples, if $E_{\rm iso}\sim10^{53.5}$erg and $d\sim50$ Mpc, using results in \autoref{fig:temp_spectrum_50Mpc}, the expected muon neutrino events within $T\simeq1$ yr for ARCA21 and IceCube can be up to $N_\nu\sim0.01$ and $\sim0.1$, respectively.

\section{Discussion} \label{sec:discussion}

We adopt a small scattering angle approximation, which basically holds if $\theta_{\rm c}\lesssim0.1$. By \autoref{eq:theta_c}, this implies that in the parameter space that is concerned, i.e., $B\lesssim1$\,nG, $d<1$\,Gpc, and/or $E_p\gtrsim10$\,EeV, the approximation can be valid. This largely leads to $\theta_{\rm j}>\theta_{\rm c}$ considered so far. However, in the case of narrow jet and strong scattering by large IGMF, the spreading of the CR beam will reduce neutrino flux significantly. 

CNPS behavior depends much on the IGMF. For large IGMF, CNPS has large duration but low neutrino flux, and vice versa. For IGMF even smaller than M1, say, $B\sim10^{-14}$G, $\tau_{\rm eff}\lesssim0.1$ s for $E_p\sim E_{\rm cut}$, CNs should be observed during the GRB bursting phase. In such case, the time-varying injection of UHECRs should be considered, and CNs may be contaminated by the UHE neutrinos produced in the GRB jet during the prompt phase \citep{2025arXiv250216946Z}. However, if the magnetic field of the source's local environment is enhanced due to the local structure, the CN arrival time may be stretched and delayed \citep[e.g.,][]{2009ApJ...690L..14M,2012ApJ...748....9T}.

CNPS behavior also depends on the source distances. The flux may be expected to roughly scale as $\propto d^{-2}$ for $d>d_{\rm eff}$. However, for low IGMF (M1 case), protons with $d_{\rm eff}(E_p)>d$ propagate through Earth in short observer times, the neutrino flux drops, and the spectrum evolves toward low energies rapidly. For larger distances, $d\gtrsim100$ Mpc, the longer propagation time will somewhat compensate for the time-integrated neutrino flux.

So far, we only consider the line of sight on jet axis. Even if off-axis, the observed CNPS flux and size is not affected, if the observer did not ``feel" the edge of the CR beam. This occurs when the observable neutrino emission region is within the CR beam, i.e., $\theta_{\rm b}-\alpha_{\rm L}>\theta_{\rm c}$, with $\alpha_{\rm L}$ the angle between the light of sight and the jet axis. However, if the GRB is observed from the vicinity of the jet edge, $\theta_{\rm j}\lesssim\alpha_{\rm L}$, it may happen that the GRB is missing but the CNPS can be observed when the CR beam spreads to the point that $\theta_{\rm b}>\alpha_{\rm L}$. 

To consider future detectability of CNPSs, in \autoref{fig:temp_spectrum}, we show the diffuse CN flux (taken from \cite{2001PhRvD..64i3010E}, which also assumes a flat proton spectrum) within a sky region of radius $0.1^\circ$ and $1^{\circ}$. The CNPS can pop up from diffuse emission in an early time, say within $\sim100$ yr. Note also that the angular resolution for future UHE neutrino telescopes, such as GRAND and IceCube-gen2, is sub-degree, $\sim0.1^\circ$. The CN sources basically appear as point sources during the pop-up period. In terms of sensitivity for IceCube-gen2 radio array and GRAND, the sensitivity for point sources estimated by the sensitivity for isotropic sources \citep{IceCube-Gen2:2020qha,GRAND:2018iaj} are similar, $\sim 3\times10^{-8} \rm GeV\, cm^{-2}s^{-1}$ for 1 yr observation.  The CNPS with fiducial values is hard to detect, but may be detected in the situation of early time, large energy, small distance and low IMGF, say, $\tau\lesssim 0.1$ yr, $f_pE_{\rm iso} \geq 10^{54.5}$erg, $d\lesssim50$ Mpc, and $B=10^{-12}$G.

If the KM3 event is induced by a nearby GRB, it is interesting to search for the historic GRB from the archive data (say, about one year before KM3 event) in the event's direction, as well as its afterglow, although the large position uncertainty is challenging to search for long-wavelength counterparts. On the other hand, the GRB may be off-axis, as discussed above, with the viewing angle $\alpha_{\rm L}\gtrsim\theta_{\rm j}$, slightly outside of the jet edge, then the KM3 event could be free of GRBs in observations.

\begin{acknowledgments}
This work is supported in China by National Key R\&D program of China under the grant 2024YFA1611402. T.-Q.H. is supported by the Special Research Assistant Funding Project of Chinese Academy of Sciences. 
\end{acknowledgments}

\appendix

\section{Effective area of neutrino telescopes}\label{sec:effective area}
We evaluate the scale factor $\xi(E_{\nu})$ using IceCube's effective area for muon track events which are prioritized in neutrino source searches.
Firstly, the effective area scales with the projected geometric cross-section of the array in the event direction.
The projection combines contributions from the top/bottom surfaces and sidewalls,
\begin{equation}
   A_{\rm proj}(\theta)=\pi r^2{\rm cos}\theta+2rh|{\rm sin}\theta|,
\end{equation}
where $r$ is the array's radius, $h$ is its height, and $\theta$ is the zenith angle of 
IceCube approximates a cylinder with $r\approx560\,{\rm m}$ and $h\approx1020\,{\rm m}$. 
ARCA21, scaled from a full block with 115 detection lines ($r\approx500\,{\rm m}$ and $h\approx648\,{\rm m}$), has a smaller $r\approx214\,{\rm m}$. 
Secondly, the effective area toward the source direction is calculated by averaging the zenith-dependent effective area, weighted by the fraction of time the source occupies each zenith angle due to Earth's rotation. 
For an IceCube-like detector at the latitude of KM3NeT-ARCA ($36^{\circ}16^{\prime}\,{\rm N}$) and with the geometry of ARCA21, the effective area for a point source at declination $\delta_s$ is:
\begin{equation}
   A_{\rm eff,\,IC\rightarrow ARCA21}(\delta_s) = 
   \frac{1}{T}\int_0^{T} A_{\rm eff, IC}\left(\theta(t|\delta_s, \varphi_{\rm ARCA})\right)\beta(\theta)dt,
\end{equation}
where $\beta=A_{\rm proj,\,ARCA21}/A_{\rm proj,\,IC}$ quantifies the projected area ratio. The up-going effective area is
\begin{equation}
   A_{\rm eff,\,IC\rightarrow ARCA21}^{\rm up} = 
   \frac{1}{1+{\rm cos(85^{\circ})}}\int_{85^{\circ}}^{\pi} A_{\rm eff,\,IC}(\theta){\rm sin}\theta d\theta.
\end{equation}
We define the scale factor as
\begin{equation} \label{eq:xi}
   \xi= \frac{A_{\rm eff,\,IC\rightarrow ARCA21}(\delta_s)}{A_{\rm eff,\,IC\rightarrow ARCA21}^{\rm up}}.
\end{equation}
This ratio spans $\xi \sim 0.7-1.2$ for neutrinos with energies from 1 PeV to 100 PeV.

\begin{figure}[thb]
   \centering
   \includegraphics[width=0.8\linewidth]{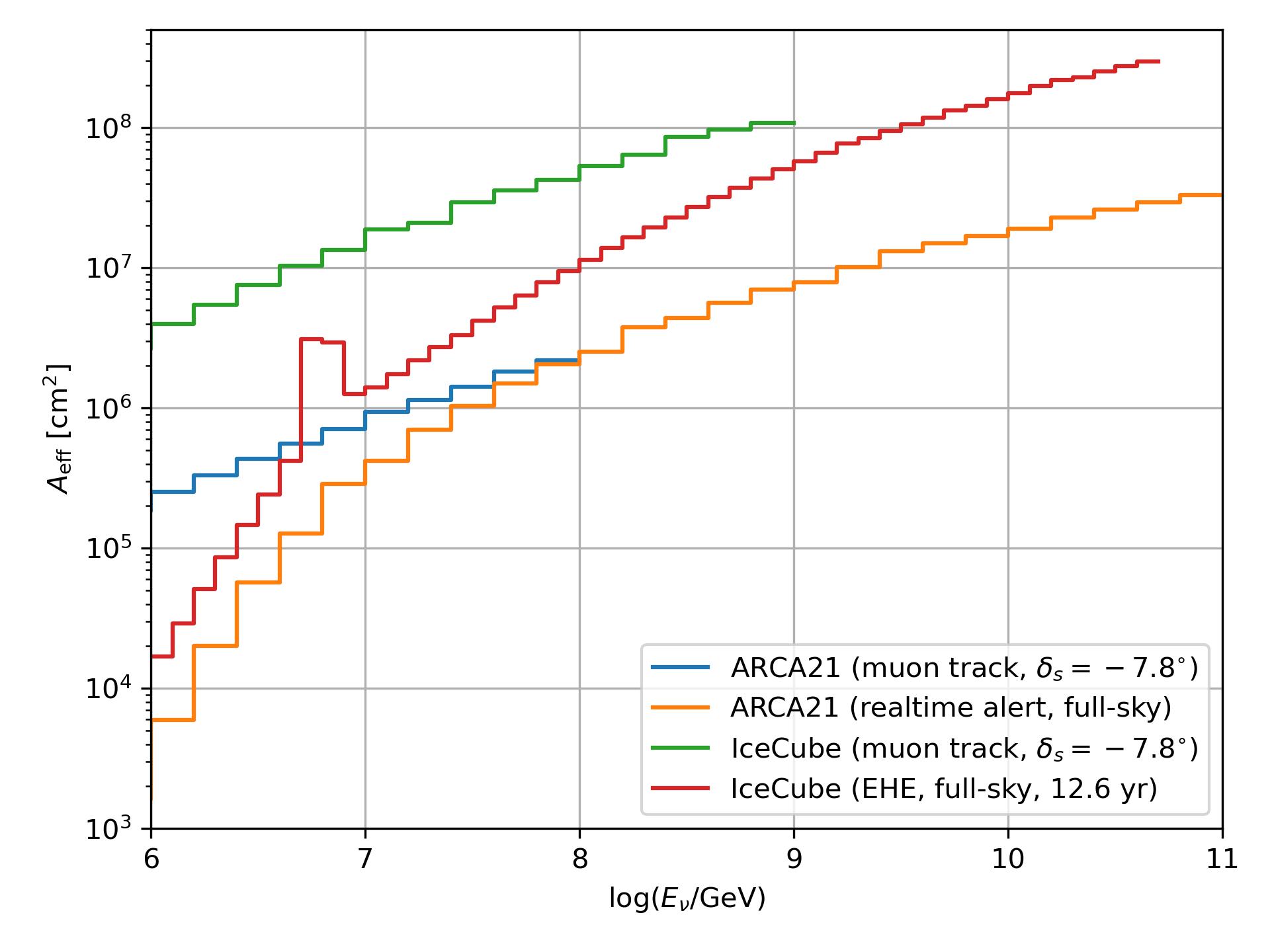}
   \caption{The effective areas of KM3NeT-ARCA and IceCube. }
   \label{fig:Aeff}
\end{figure}

\bibliography{sample7}{}
\bibliographystyle{aasjournalv7}


\end{CJK*}
\end{document}